\documentclass[11pt]{article}
\usepackage{hyperref}
\hypersetup{colorlinks=true, urlcolor=blue,citecolor=blue}
\usepackage[usenames]{xcolor}
\makeatletter
\@addtoreset{equation}{section}
\makeatother

 \def\bd{\begin{document}} \def\ed{\end{document}}
\def\ds{\documentstyle} \let\fr=\frac \let\bl=\bigl \let\br=\bigr
\let\Br=\Bigr \let\Bl=\Bigl
\let\bm=\bibitem
\let\na=\nabla
\let\pa=\partial \let\ov=\overline
\newcommand{\be}{\begin{equation}}
\newcommand{\ee}{\end{equation}}
\def\ba{\begin{array}}
\def\ea{\end{array}}
\newcommand{\ho}[1]{$\, ^{#1}$}
\newcommand{\hoch}[1]{$\, ^{#1}$}
\newcommand{\bea}{\begin{eqnarray}}
\newcommand{\eea}{\end{eqnarray}}
\newcommand{\ra}{\rightarrow}
\newcommand{\lra}{\longrightarrow}
\newcommand{\Lra}{\Leftrightarrow}
\newcommand{\ap}{\alpha^\prime}
\newcommand{\bp}{\tilde \beta^\prime}
\newcommand{\tr}{{\rm tr} }
\newcommand{\Tr}{{\rm Tr} }
\newcommand{\NP}{Nucl. Phys. }
\newcommand{\tamphys}{\it
The Blackett Laboratory, Imperial College London,\\ Prince Consort Road, London SW7 2AZ}

\newcommand{\auth}{M. J. Duff\footnote{m.duff@imperial.ac.uk}}

\begin{document}
\hfill{}

\hfill{Imperial/TP/2011/mjd/4}

\vspace{24pt}

\begin{center}
{ \large {\bf String and M-theory: answering the critics\footnote{Contribution to the Special Issue of Foundations of Physics: ``Forty Years Of String Theory: Reflecting On the Foundations'', edited by Gerard 't Hooft, Erik Verlinde, Dennis Dieks and Sebastian de Haro.}
}}

\vspace{24pt}

\auth

\vspace{10pt}

{\tamphys}

\vspace{24pt}

\underline{ABSTRACT}

\end{center}

Using as a springboard a three-way debate between theoretical physicist Lee Smolin, philosopher of science Nancy Cartwright and myself, I address in layman's terms the issues of why we need a unified theory of the fundamental interactions and why, in my opinion, string and M-theory currently offer the best hope. The focus will be on responding more generally to the various criticisms. I also describe the diverse application of string/M-theory techniques to other branches of physics and mathematics which render the whole enterprise worthwhile whether or not  ``a theory of everything'' is forthcoming.

\vfill
\leftline{}

\newpage

\section{String and M-theory}
\label{intro}
When the editors of this Special Issue on String Theory invited me to contribute, they informed me that ``it aims at a status review of string theory where recent criticism--such as not standing up to the claims it made in the past and not making contact with the real world--is discussed as well''.  I think my invitation was partly motivated by a three-way debate between theoretical physicist Lee Smolin, philosopher of science Nancy Cartwright and myself that took place at The Royal Society for the Arts in London in 2007, centred on Lee Smolin's critique of string theory ``The Trouble with Physics'' \cite{Smolin}, which I recall in  Section \ref{debate}. Here I use this debate as a starting point to answer more generally the critics of string theory and its successor M-theory. Although many problems remain to be solved, I believe that much of this criticism is misguided or misinformed\footnote{Another book attacking string theory appeared at about the same time: ``Not even wrong'' \cite{Woit1} by Peter Woit, Computer Administrator and Senior Lecturer in Discipline at Columbia University.}. I will argue in favour of the need for a unified theory of the fundamental interactions and explain why, in my opinion, M-theory currently offers the best hope notwithstanding the lack of experimental confirmation.  I invoke some historical precedents in Section \ref{theory} to support the view  that rapid confirmation by experiment is a poor guide to the eventual success of theories of physics.  Section \ref{maverick} refutes the conspiracy-theory that there is a theory-conspiracy by the string  ``establishment'' to suppress rival approaches. I will also mention in Section \ref{repurpose} the contributions of string and M-theory to pure mathematics and recent applications of string and M-theory techniques to other branches of physics, such as cosmic strings, quark-gluon plasmas, fluid mechanics, high-temperature superconductors and quantum information theory.  These largely serendipitous discoveries represent useful progress whether or not M-theory turns out to be a ``theory of everything'', but would never have happened had the stop-string-theory critics had their way. Finally, in Section \ref{matter},  I maintain that, misinformed though some criticisms of string and M-theory may be, they can still be very damaging and so require a response. In this respect I differ from some of my colleagues.

 The article is aimed at non-specialists, not only because public understanding of science is a good thing, but also because decisions about the the future direction of scientific research are increasingly being made by non-scientists\footnote{Here in the UK, commentators expressed delight at the news that the Treasury, up until recently the only government department not to have a chief scientific advisor, has finally appointed one.  Few seemed bothered that his PhD is in economics.}, some of whom are hostile to string theory. 

How did the universe begin? What are its fundamental constituents?  What are the laws of nature that govern these constituents?  The situation we faced at the end of the 20th century, notwithstanding the great success of the standard model of particle physics and the standard cosmological model in providing partial answers to these questions, was that the two main pillars of 20th-century physics, quantum mechanics and Einstein's general theory of relativity, seemed mutually incompatible.  Quantum theory deals with the very small: atoms, subatomic particles and the forces between them. General relativity deals with the very large: stars, galaxies and gravity, the driving force of the cosmos as a whole. The dilemma is that on the microscopic scale, Einstein's theory fails to comply with the quantum rules that govern the behaviour of the elementary particles, while on the macroscopic scale black holes are threatening the very foundations of quantum mechanics. Something big has to give.  This augurs a new scientific revolution.

But no one said it was going to be easy. Constructing an all-embracing theory that unites gravity and quantum mechanics and goes on to describe all physical phenomena is probably the most ambitious challenge in the history of science. Finding ways in which it can be tested empirically will then be equally difficult, given that the gravitational force is 40 orders of magnitude weaker than the other forces of nature.

Many physicists nevertheless believe that this revolution is already under way with the theory of superstrings. As their name suggests superstrings are one-dimensional string-like objects. Just like violin strings, these relativistic strings can vibrate and each mode of vibration, each note if you like, corresponds to a different elementary particle. This note is a quark, this one a Higgs boson, that one a graviton and so on.  One strange feature of superstrings is that they live in a universe with nine space dimensions and one time dimension. Since the world around us seem to have only three space dimensions, the extra six would have to be curled up to an unobservably small size (or else rendered invisible in some other way) if the theory is to be at all realistic. Fortunately, the equations admit solutions where this actually happens.

The main reason why theorists are so enamoured with string theory is that it seems at last to provide the long-dreamed-of consistent quantum theory of gravity and holds promise incorporating and extending the standard models of particle physics and cosmology.  

String theorists are the first to admit that the theory is by no means complete but is constantly undergoing improvement in the light of new discoveries. For example, one of the problems with the form of the theory developed in the 1980s was that there was not one but five mathematically consistent superstrings. If one is looking for a unique theory of everything, five theories of everything seem like an embarrassment of riches.

In 1995 the theory underwent a revolution when it was realized that these five strings were not after all different theories but just five corners of a deeper and more profound new theory, called ``M-theory'' \cite{Duff1998}.  M-theory involves membrane-like extended objects \cite{Sutton} with two space dimensions and five space dimensions that themselves live in a universe with eleven spacetime dimensions (ten space and one time). For the purposes of this article I will often use the words ``string theory'' but they are meant to imply M-theory as well.

String and M-theory continue to make remarkable theoretical progress, for example by providing the first microscopic derivation of the black hole entropy formula first proposed by Hawking in the mid 1970s. Solving long outstanding theoretical problems such a this indicates that we are on the right track.

But, as often happens in science, M-theory presented new problems of its own, not least of which is that its equations admit even more different solutions than string theory does and at the moment we have no idea which one, if any, nature should pick to describe our universe. Theorists are divided on this issue. Some think that when we understand the theory better, we will understand why one unique universe will be singled out, thus answering in the negative Einstein's question ``Did God have any choice in creating the universe?''
Others think that there are indeed many, possibly infinitely many, different universes and we just happen to be living in one of them.  The problem of how to choose one universe out of a large number of mathematically possible universes is sometimes called the ``landscape'' problem. Some string theorists like Lenny Susskind regard the landscape as a virtue to be exploited, as it fits in with ideas of the ``multiverse'' advocated by astronomers like Martin Rees; others like David Gross are sceptical.

Lacking an answer to this problem, the theory is as yet unable to give a definitive ``smoking-gun'' experimental prediction that would render the theory falsifiable.  Of course there are some generic features of string theory like supersymmetric particles, extra dimensions, microscopic black holes and cosmic strings. If we are lucky, some of these may be confirmed at the Large Hadron Collider in Geneva, or perhaps in the next generation of astrophysical observations.  But I am doubtful whether the kind of issues we are considering here will be resolved any time soon.

Some critics say that since you have been working on M-theory since 1995, it is time to give up. Others like myself say that it is still early days. There is an important point here that is rarely mentioned by critics of string theory but which I would like to stress. Let us suppose for the sake of argument that everyone were to abandon string theory tomorrow, sharpen their pencils and ask ``Where do we go from here?''. The landscape problem would not have gone away. The problem of how to choose one physical universe out of a large number of mathematically possible universes is a problem that any attempt to provide a final theory is going to have confront. Why do we appear to live in just four dimensions? Why is the number of fundamental forces the four of gravitational, electromagnetic, strong and weak nuclear? Why are there just three families of quarks and leptons? These riddles are not unique to string theory and at the moment none of the alternative theories has any answers to them.

\section{The debate}
\label{debate}
\subsection{Response to Lee Smolin}

Who can dispute that the ultimate goal of a scientific theory is to make experimentally testable predictions? Who will challenge the need to keep an open mind and listen to unorthodox views?  Who can disagree with the assertion that our current understanding is only partial and that the ultimate truth has yet to be uncovered? What Lee Smolin said in the London debate \cite{Smolinrsa} was so uncontroversial that, had I confined my response \cite{Duffrsa} to these remarks, the evening would have would have fizzled out in a bland exchange of truisms\footnote{In the version of events appearing on Peter Woit's blog \cite{Woit2}: ``Smolin sat down. Duff stood up. It got nasty.  The trouble with physics, Duff began, is with people like Smolin''. Just to set the record straight,  what I actually said was: ``The trouble with physics, Ladies and Gentlemen, is that there is not one Lee Smolin but two. On the one hand, there is the reasonable Lee Smolin who has just spoken to you, and with whom I hope to share a drink later this evening.....Unfortunately for physics...there is the Lee Smolin who wrote this book ..And this Lee Smolin is far from reasonable''. When someone who had read the transcript pointed this out, Woit responded:
``It does not purport to be a direct quote of what Duff said. It is a direct quote of how that person chose to characterize what Duff said. It reflects precisely how that person interpreted what he was hearing Duff say.''}.  I decided therefore to focus on his book ``The Trouble with Physics'', which was after all the topic of the debate, and here the gloves were off. For example, the cover asserts:

\begin{itemize}
\item{}
`` As a scientific theory, string theory has been a colossal failure and is dragging down the rest of physics''  
\end{itemize}

And that is just for starters. The book itself has many similarly unpleasant, but in my opinion similarly inaccurate, things to say both about string theory as a science and string theorists as scientists.  Moreover, such accusations had by then received quite a wide publicity. For example, in the previous week's Mail on Sunday \cite{Ritchie} a review of the book concluded:

\begin{itemize}
\item
``String theory has made no discernable progress after 20 years''
\item
``Just a huge amount of unprovable conjecture''
\item
``Of the hundreds of research appointments in American Universities since 1990, only three have gone to non-stringies''
\end{itemize}

Now I don't mind that journalists are ignorant about string theory; most people are. But is it too much to ask that they check the facts? First, as described above, the ``no progress'' claim is simply silly. The ``unprovable conjecture'' claim is also false. As discussed in Section \ref{theory}, conjecture and refutation is a time-honoured way to proceed in scientific research. String theory is no different. Sometimes our conjectures are proved, sometimes disproved, sometimes no proof is forthcoming but such a vast amount of circumstantial evidence is accumulated that we adopt them as a working hypotheses until the proof is eventually nailed down. How does The Mail on Sunday know that that the conjectures so far unproved are unprovable?  Perhaps they would care to lend me their crystal ball.

Finally, a couple of phone-calls would have easily refuted the ``only three appointments'' claim.  I spent the years 1999-2005 at the University of Michigan and the last two permanent faculty appointments in theoretical physics were both to non-stringies. Multiply that by the dozens of similar departments in the US and you see what a flagrant misrepresentation the ``three out of hundreds'' really is. 
I might add that it is equally untrue in the UK. At my own department at Imperial College, for example, 7 out of the 12 faculty appointments since 1990 were to non-stringies. Of course, the Mail on Sunday has never been a paper to allow facts to get in the way of a good story. 

Where I do agree with their review (and with most other reviews I have read) is that the book is a venomous attack on string theory and its practitioners. Now don't get me wrong. If Lee had confined his sociological criticisms to saying that some string theorists are arrogant, exclusive and unwilling to listen to unorthodox views, then that is fair game. There is probably some truth in that. See Section \ref{maverick}. But to conflate the sociology with science by accusing them of being bad physicists is altogether different.

Nor is gullibility confined to journalists. Philosopher A. C. Grayling \cite{Grayling} accepts uncritically everything in the book, proclaiming that rival approaches
\begin{itemize}
\item{}
``...include such theories as loop quantum gravity, doubly special relativity and modified Newtonian dynamics. All of these make testable predictions, and if wrong can be shown to be so, itself always an advance in science; and therefore, unlike string theory, they are `genuine scientific theories' ''.
\end{itemize}
Stephen Hawking recently complained that ``philosophy is dead'' because philosophers have failed keep up with scientific developments. If the above pronouncement of A. C. Grayling is any guide, one can only agree.

Over and over again, Smolin uses the stick of ``no falsifiable predictions'' with which to beat string theory.  I have read his book of suggested alternatives from cover to cover and guess what? There is not a single falsifiable prediction!  Instead we are told\footnote{The page numbers of ``The Trouble with Physics'' in the text refer to the version I was sent to review by Nature Physics and may be different in published versions.}
:

\begin{itemize}

\item{P 185} ``There are HINTS of new experimental discoveries''

\item{P 232} `` Loop quantum gravity PROMISES to be able to make a sure prediction''

\item{P 232} ``Recent results SUGGEST that this is exactly the right particle physics, the standard model''

\item{P 234} ``It is TOO EARLY  to tell if it works well enough to give unambiguous predictions''

\item{P 235} ``Rather, it SEEMS to lead to a unique theory, which will either be in agreement with experiment or not ''
\end{itemize}

Shorn of the rhetoric, he is simply saying ``I have a theory.  I don't yet know whether it is right but I am optimistic that I will eventually be able to make a falsifiable prediction that will be supported by experiment'', which is exactly what the string theorists are saying.  Who has the more justification for these hopes? Well, time will tell, but most of the bright young people, as Smolin himself concedes, are voting with their feet and opting for string theory.

In the Chapter ``Seers versus Craftspeople'' Smolin divides scientists into two camps: There are the ``craftspeople'' who though competent at doing tedious calculations are merely unimaginative drones doing what others have told them to do, and then there are the ``seers'' who have all the real vision and creative genius.  Needless to say, Smolin consigns the entire String community to the dustbin of craftspeople:
\begin{itemize}
\item{P 287} `` Why are strings theorists not seers?''

\item{P 290} ``I can think of no mainstream string theorist who has proposed an original idea about the foundations of quantum theory or the nature of time.''

\item{P 287}  "Nor does it take much foresight or courage to think about these things when hundreds of other people are thinking the same thoughts.''
\end{itemize}
So when Nobel Laureates Murray Gell-Mann, Abdus Salam, Steven Weinberg and David Gross, all gave up what they were doing in order to study String Theory, when Fields Medalists  Michael Atiyah and Edward Witten established deep connections between pure mathematics and string theory and when Stephen Hawking devoted his latest book to M-theory, they were all, according to Smolin, lacking foresight: 
\begin{itemize}
\item{P 289} ``so little fulfillment is exactly what you get when a lot of highly trained master craftspeople try to do the work of a seer''
\end{itemize}
Who then are the seers? Well, Smolin says\footnote{In the debate I quoted from the version of the book I was sent to review by the journal Nature Physics:  ``I think of myself as a seer'', but this did not appear in the published version. My apologies to Lee.}
\begin{itemize}
\item{P 287}  ``I think of myself as a would-be seer''
\end{itemize}
although it is left unclear how and when full seerdom is attained.

To make matters worse, the end of his book is devoted to condemning racism and sexism in Higher Education. When all this is mixed up with attacks on string theory, the casual reader is left with the impression that practically all the ills of society are to be laid at the feet of string theorists.

How should one react to such a book? My own reaction is best summed up by an analogy with football (soccer). Let us suppose that the manager of a football club, say Watford, were to write a book called ``The Trouble with Football'' whose cover announces that another club, let us say Manchester United, are, as a football team `` a colossal failure'' who are ``dragging down the rest of football''.

Why? Because although those United players may be scoring all the goals, winning all the trophies and garnering generous sponsorship, they are merely super-fit craftspeople. All the players with real vision and creative genius, the seers, are playing for Watford! And the reason that most of the up and coming young talent wants to play for United rather than Watford is because they are deluded.  

How would such a book be received? Would it persuade discerning football fans? I do not think so, for the obvious reason. They would say that the only way to get more people to support your club and fewer to support your rivals, and the only way to persuade young talented players to join your team, is not by writing books but by playing the game.  

And so it is in Theoretical Physics. The battle for the correct theory will not be won on the bookshelves of Barnes and Noble, nor in the debating chamber of the Royal Society for the Arts. The battle will be won in the pages of the scholarly scientific journals or, in their modern guise, on the electronic archives on the internet (\href{http://arxiv.org/}{{\tt {\color{blue}http://arxiv.org/}}}). The way to persuade your scientific colleagues that you have a good theory that is worthy of support is to publish your findings and make the most convincing case you can.

Sadly, many critics of string theory, having lost their case in the court of Science are now trying to win it in the court of Popular Opinion.  Of course you may well ask,  if the majority of theoretical physicists were in favour of string theory, would it really matter if the general public or scientists in other fields were not? In Section \ref{matter}, I will argue that it does.

\subsection{Response to Nancy Cartwright}

In her presentation \cite{Cartwrightrsa}, philosopher Nancy Cartwright opposed not only string theory but all attempts at unification:
\begin{itemize}
\item{}
``Now according to Smolin, string 
theorists often dissent: they claim to know string 
theory is best. Indeed more than one of them 
(this I know is true) more than once have 
prophesised that the end of physics is in view 
with string theory. [As if to say]: ``Now we don't need anymore 
experiments in living. We know how to live. 
All we have to do is polish off a few details.''
Now in ordinary life, we have a name 
for this kind of attitude -- hubris -- and most 
of us dislike it. Indeed, distrust it. 
Philosophers of science have a complicated 
name for a very simple argument as to why it 
is a bad idea in physics. The argument is 
called ``The pessimistic meta-induction''. And 
it's this. Our all best, most successful, most 
lovely basic theories in the past have been 
shown to be wrong. Any good method of 
reasoning from past to future tells us that our 
current best, most successful, most lovely 
basic theory will be wrong as well. ''
\end{itemize}

\begin{itemize}
\item{}
``We now take most of our great theories of the past to be  
fundamentally mistaken. But we touted them as
great, just because they did so well in all the 
usual epistemic virtues. So epistemic virtue has 
shown itself not to be a good guide to truth! 
We are then without a reliable, well 
grounded substantive theory of theory-choice. ''
\end{itemize}

Leaving aside her conflation of good science and good manners (few of the Nobel laureates I have met count lack of hubris amongst their epistemic virtues), the assertion that all our best theories in the past have now been shown to be wrong is, is my opinion, a funny way of putting it. Take Newtonian mechanics, for example. If I used it to predict where a cannonball will land, would Nancy Cartwright be willing to stand there? If not,  I assume her meaning of ``wrong'' is that Newton's theory has limited regimes of validity and outside those regimes must be replaced by something better, such as Einstein's general relativity which subsumes both Newtonian mechanics and special relativity (and is now used in GPS devices, where it gives the ``right'' answers). By the same logic, however, the failure of Einstein's theory in the subatomic regime calls for its replacement by something better that subsumes both general relativity and quantum mechanics (a theory, by the way, that so far has been ``right'' every time). In my view, this historical trend does not undermine unification; it celebrates it.

I am glad that no philosophers of science prevented Newton from unifying terrestrial and celestial gravity, or Faraday and Maxwell from unifying electricity and magnetism or dissuaded Glashow, Salam and Weinberg from unifying electromagnetism and the weak nuclear force.
\footnote{I do not share Lubos Motl's extreme views on politics, global warming, and sometimes not even string theory. However, he occasionally has some good physics summaries, including a recent one giving a nice history of the triumphs of unification\cite{Motl}.}.

Cartwright's anti-unification argument is based on ``pessimistic meta-induction'':
\begin{itemize}
\item{}

`` In his [Smolin's] book, 
the unification is the name of the game. But 
why? It is in no better state than any of the 
other so-called epistemic virtues when it 
comes to claims to truth conduciveness. 
Pessimistic meta-induction yet again, we have 
zillions of failed attempts at unifications. Not 
the ones that Lee's interested in but when I 
studied physics, worked on a lot more 
practical things like quantum optics to see 
how lasers worked. There were thousands of 
failed attempts at unifications. And hosts of 
previously successful unifications we have 
now discarded. So pessimistic meta-induction 
shows that in the past unification hasn't been 
truth conducive; no reason to think it is 
particularly in the future. And a handful of 
unifications that we still accept do not much 
counter the weight for making the inductive 
argument. ''
\end{itemize}

It is difficult to resist yet another sporting analogy. In the game of football, most attacks on goal end in failure, but every now and then there is a breakthrough and somebody scores. If I understand it correctly,  the philosophy of ``pessimistic meta-induction'' says that if your team has not scored by half-time, you should give up and go home.

\section{Theory and experiment}
\label{theory}

This is perhaps a good time to pause and ask about the relationship between theory and experiment in physics.

The image held by many members of the public and, sadly, by some science journalists, is as follows. Some theoretical genius thinks up a bright idea while sitting in an armchair and shouts ``Eureka!'' The next day he or she writes a paper summarizing the theoretical ideas and suggesting ways of testing them in the laboratory.  Shortly afterwards, an experimental colleague performs the experiment and confirms the theory. The following December both go to Stockholm to collect their Nobel prizes. 

Alas, the reality is usually completely different. Theories rarely spring fully formed from the minds of their discoverers.  They usually begin with the germ of an idea that needs to be nurtured and nourished and then gradually improved upon by others.  Let us consider some examples from the history of physics.
\begin{itemize}
\item{Black holes}

The existence of black holes, objects so dense that not even light can escape, is now well supported by astronomical observation, particularly from X-ray emission from binary stars and active galactic nuclei. But the theory of black holes goes back to John Mitchell in 1784 and Pierre Laplace in 1786. They reasoned that there was a critical radius for a massive body below which the escape velocity would exceed the velocity of light. Though basically correct, these ideas did not reach theoretical maturity, however, until the arrival of Einstein's general theory of relativity in 1916 and the discovery of what we now call the Schwarzschild black hole solution of the Einstein equations. They were still not taken seriously, even by Einstein.  In 1939 Oppenheimer and Snyder showed that after burning up their nuclear fuel, sufficiently massive stars would undergo gravitational collapse and form black holes. But even this work was largely ignored until the observational evidence for black holes started to accumulate in the 1970s. 

So I find myself in disagreement with Smolin when he says: 
\begin{itemize}
\item{}  ``...there is no precedent in the history of science, since at least the late eighteenth century, for a proposed major theory going more than a decade before either failing or accumulating impressive experimental or theoretical support''
\end{itemize}

\item{Quantum entanglement}

   One of the strangest and most fascinating physical phenomena is that of quantum entanglement, where a measurement by Alice can affect a measurement by Bob
no matter how far apart they are.  The basic problem was set forth in 1935 by Einstein, Podolsky and Rosen who pointed out that there was a conflict between quantum theory and local reality, or what Einstein called ``Spooky action at a distance".  But it was not until 1964 that CERN theorist John Bell came up with an empirically falsifiable prediction. And it was not until 1982 that experimentalist Alain Aspect was able to conduct Bell's suggested experiment. He verified empirically that quantum mechanics is right, Einstein was wrong, and local realism has to go out of the window. 

So once again I disagree with Smolin when he says:
\begin{itemize}
\item{} ``Before the 1970s, theory and experiment had developed hand-in-hand. New ideas were tested within a few years, ten at most. '' 
\end{itemize}

\item{Atoms}

The idea that everything is made of tiny atoms is usually attributed to the Greek thinker Democritus round about 400BC. It gained more support in the 19th century but it was not until the advent of quantum mechanics at the start of the 20th century that the atomic theory came of age. 
\end{itemize}

There are lots of other examples in the history of science where good theoretical ideas waited a long time for experiment. Bose-Einstein condensation was predicted in 1925 but not confirmed until 1995; Yang and Mills invented non-abelian gauge bosons in 1954 but the W-boson was not discovered until 1982. Yet Smolin says:
\begin{itemize}
\item{P 236} ``When you look back at the history of physics, one thing sticks out: When the right theory is finally proposed, it triumphs quickly.''
\end{itemize}

There are two morals to be drawn from these tales. The first more obvious one is that we frequently have to wait for the technology to catch up with the theory. Many theoretical predictions were not confirmed until the arrival of more sophisticated hardware, more powerful telescopes or particle accelerators. The second moral, harder for the layman to appreciate but the one of vital importance for this debate, is that it is frequently takes a long time for an original theoretical idea to mature to the stage where it can be cast into a smoking-gun prediction that can be tested experimentally. 

Many confused critics of string theory have declared that unless you are making falsifiable predictions you are not doing science. Were Democritus, Mitchell, Laplace, Einstein, Podolosky, Rosen, Oppenheimeer and Snyder not therefore doing science?  Of course they were! They were simply ahead of their time.

If we are to reject the 1995 M-theory because its ten years without definitive empirical proof are already up, then we must also reject gravitational waves (1916), the cosmological constant (1917), extra dimensions (1926), the Higgs boson (1964), supersymmetry (1971) and, er, Lee Smolin's favourite theory, loop quantum gravity (1986).

The job of theoretical physicists is twofold: first, to explain what our experimental colleagues have discovered; and second, to predict phenomena that have not yet been found. The history of scientific discovery shows that progress is achieved using both methods.

Quantum theory, for example, was largely driven by empirical results, whereas Einstein's general theory of relativity was a product of speculation and thought experiments, as well as advanced mathematics.

Speculation, then, is a vital part of the scientific process. When Paul Dirac wrote down his equation describing how quantum particles behave when they travel close to the speed of light, he wasn't just explaining the electron, whose properties had been well established in experiments. His equation also predicted the hitherto undreamed-of positron, and hence the whole concept of antimatter.

Such speculation is not a flight of fancy. It is always constrained by the straightjacket of mathematical consistency and compatibility with established laws. Even before it was tested experimentally, Einstein's theory of general relativity had to pass several theoretical tests. It had to yield special relativity and Newtonian mechanics in those areas where they were valid, as well as predict new phenomena in those where they were not.

It is a common fallacy that physics is only about what has already been confirmed in experiments. Commentators  in New Scientist have unfairly compared the study of cosmic strings - macroscopic objects that may have been formed in the early universe - to UFOs and homeopathy, on the grounds that cosmic strings have yet to be observed \cite{hanlon}. Another stated that until M-theory is backed by empirical evidence, it is no better than ``faith" \cite{ed}. To its credit, New Scientist did publish my objections \cite{Duffns}. 

Yet support for superstrings and M-theory is based on their ability to absorb quantum mechanics and general relativity, to unify them in a mathematically rigorous fashion, and to suggest ways of accommodating and extending the standard models of particle physics and cosmology. No religion does that.

By the same token, some alternative ideas purporting to be theories of everything have had to be rejected even before their predictions could be tested - not on the grounds of faith but because they were mathematically erroneous. See Section \ref{maverick}.  What separates theoretical speculation from faith is that we modify or reject theories in the light of new evidence and discovery.

The most effective way for critics of M-theory to win their case would be to come up with a better alternative. So far nobody has.

\section{The mavericks, the loners, ...}
\label{maverick}

In his review \cite{Grayling} of ``The trouble with physics'', philosopher A. C. Grayling goes on to say:
\begin{itemize}
\item{}

``... finally, but by no means least, he [Smolin] enters an eloquent plea on behalf of the mavericks, the loners, the original thinkers, the sceptics, the unusual and eccentric minds, who he believes are needed to free theoretical physics from the impasse it currently finds itself in''
\end{itemize}
It is really the case that there are brilliant loners out there and that there is some kind of conspiracy by the physics ``establishment'' to prevent their voices being heard?

Anyone who has read my own critiques of superstring orthodoxy \cite{Duff:1987dk,Duff:2001jp}
will know I share the concern that, especially in the United States, string theory is not immune from the pressure to conform (though I do not consider it is worse than other subjects in this respect).  So I am all in favour of mavericks, but with one important proviso: I am favour of mavericks whose unorthodox ideas turn out to be right.  Let me explain.

Of all the intellectual disciplines, theoretical physics is one of the toughest in which to gain expertise, requiring years of dedicated study and a firm grasp of higher mathematics. There are no shortcuts. It is a curious paradox, therefore, that it also attracts an army of unqualified wannabes and crackpots. Like many of my colleagues in theoretical physics, every few days I receive an email or a letter or a phone call from someone claiming  to have proved X wrong (where X= Einstein, Bohr, Dirac, ....),  promoting their own theories of everything and taking a pot-shot at mainstream research, especially string theory. It may be a truism that all great ideas in science seemed crazy at the time. Unfortunately, the converse is not true, so it's a case of hitting the delete key. But is there not the danger, you say, that some undiscovered genius  will be overlooked?  Well, all I can say is that, in the forty years I have been doing research, it hasn't happened yet.

The internet, where everyone is an expert, now provides their ideal forum. Attempts at sensible commentary or discussion on the blogosphere are usually quickly overwhelmed by a cacophony (the collective noun?) of crackpots.

Unfortunately, journalists (and, it seems, philosophers) love them! Their favourite story line is that of the outsider who ``turns established physics upside-down''. Senior physicists are ``stunned'' (whereas in fact the only words of support usually come from other mavericks anxious to air their own outlandish proposals.) The story is not complete without some reference to an establishment conspiracy to censor the offending discovery. I once wrote to the editor of New Scientist to suggest they do a ``where-are-they-now?'' issue that tracks the inevitable descent into oblivion of these ``pioneers'', following their fifteen minutes of fame in his magazine. He politely declined.

A recent cause celebre along these lines was provided by Garrett Lisi, who wrote a paper entitled ``An exceptionally simple theory of everything'' \cite{Lisi}, which claims in its opening sentence to unify ``all fields of the standard model and gravity''.  Lisi is by no means a crackpot but he holds no university position and his paper has not been published in a peer-reviewed journal.  So when Lee Smolin described him as the next Einstein, the publicity juggernaut moved into overdrive:

\begin{itemize}
\item{The Daily Telegraph \cite{Highfield1}:}
Surfer dude stuns physicists with theory of everything:

``An impoverished surfer has drawn up a new theory of the universe, seen by some as the Holy Grail of physics, which has received rave reviews from scientists.

Although the work of 39 year old Garrett Lisi still has a way to go to convince the establishment, let alone match the achievements of Albert Einstein, the two do have one thing in common: Einstein also began his great adventure in theoretical physics while outside the mainstream scientific establishment, working as a patent officer, though failed to achieve the Holy Grail, an overarching explanation to unite all the particles and forces of the cosmos.
Now Lisi, currently in Nevada, has come up with a proposal to do this. Lee Smolin at the Perimeter Institute for Theoretical Physics in Waterloo, Ontario, Canada, describes Lisi's work as ``fabulous''. ``It is one of the most compelling unification models I've seen in many, many years'' he says.

So far, all the interactions predicted by the complex geometrical relationships inside E8 match with observations in the real world.``How cool is that?" he [Lisi] says.''
\end{itemize}

\begin{itemize}
\item{The Economist \cite{Economist}:} Geometry is all. A shape could describe the cosmos and all it contains. 

``..the theory has several appealing facets. It is elegant. It is expected to make testable predictions. Unlike some of the more complicated efforts to devise a theory of everything, this one should either succeed relatively rapidly or fail spectacularly. And that is more than can be said for three decades of work by other physicists.''
 \end{itemize}
 
 Even respectable journals such as Physics World, the official publication of The Institute of Physics here in the UK, devoted its July 2008 front cover to Lisi's paper together with an accompanying article and editorial. 

There was just one small problem: Lisi's paper was incorrect.

By incorrect, I don't just mean it did not fit in with current thinking; I mean it had mathematical errors concerning the group E8 that rendered the claim in his opening sentence to unify ``all fields of the standard model and gravity'' just plain wrong.  Nature (and the standard model of particle physics) has three chiral families of quarks and leptons.``Chiral'' means they distinguish between left and right, as they must to account for such asymmetry in the weak nuclear force. But as rigourously proved by Jacques Distler and Skip Garibaldi, Lisi's construction permits only one non-chiral family. Their paper, unlike Lisi's, appeared in a highly-respected peer-reviewed journal \cite{Distler}.  How cool is that?

So did Lisi withdraw or modify his paper? Did the Daily Telegraph, the Economist, New Scientist, Physics World apologise for misleading their readers? Did Lee Smolin have second thoughts about the next Einstein? On the contrary, the juggernaut just kept rolling along!.

Physics World, for example, never mentioned the Distler-Garibaldi paper, but in yet another (fifth?) article on Lisi, this one devoted to his (presumably not-peer-reviewed) T-shirt, stated  

\begin{itemize}
\item{Physics World \cite{Johnston}:}

``Lisi's theory has received a mixed response, with some leading physicists including the Perimeter Institute's Lee Smolin praising it while others like Jacques Distler of the University of Texas find fault with it.''
 \end{itemize}
 Now a mathematical theorem is, or should be, the least controversial thing in the world. It is not merely uncontroversial; it is incontrovertible.  But in the land of science journalism, opinions trump theorems. A counting of the of web links in the Physics World article gives the score: Lisi 5, Distler-Garibaldi 0. 
 
 I wrote to Physics World, pointing this out. They responded with yet another Lisi article \cite{Durrani} linking his work to an experimental paper in condensed matter physics claiming to have spotted E8 patterns in a solid material for the first time. Physics World failed to say that the E8 effect reported by the experimentalists was first predicted in 1988 by Alexander Zamolodchikov (who is, by the way, a string theorist).  The discoveries of E8 in eleven-dimensional supergravity in 1979, of E8 xE8 in string theory in 1984 and more recent (peer-reviewed) papers on E8, E9, E10, E11 also go unmentioned.

Scientific American carried an article by Lisi and Wetherall repeating several of the discredited claims  \cite{Lisisciam}.  The Daily Telegraph even said that Lisi, though wrong, was nevertheless a role model:
\begin{itemize}
\item{The Daily Telegraph \cite{Highfield2}:}
Surfer dude's theory of everything: the magic of Garrett Lisi.

``Much of the excitement was because Lisi's theory seemed to challenge string theory the dominant contender for a "theory of everything", over which there has been a bitter intellectual war. Its proponents (mostly superstar theorists) argue that the theory, which relies on tiny, subatomic ``strings" vibrating across multiple dimensions, is too beautiful to be ignored.

Yet even if Lisi is wrong--as is usually the case with attempts to erect such all-encompassing theories-- the world needs more surfer dudes. Lisi's effort captured the public imagination because, though the then 39-year-old had a doctorate, he did not work in the establishment, but was backed by a little money from a privately funded research institute called FQXi. We need more independent spirits like him, and others outside the mainstream, such as James Lovelock, the maverick environmentalist.''
\end{itemize}

In the The New Yorker \cite{ny} he was right and wrong at the same time:
\begin{itemize}
\item{The New Yorker, start of the article:}

`` It took Lisi six months to work through the idea. He matched every photon, gluon, weak boson, graviton, Higgs boson, quark, neutrino, and electron, in each of the first, second, and third generations--every particle in the standard model--to symmetries of E8.''

\item{The New Yorker, later in the article:}

``His theory is still incomplete--missing, for example, a satisfying link to the second and third generations. ''

\item{The New Yorker, later in the article:}

``Not only can one never hope to get 3 generations out of this Theory of Everything, it appears that one can't even get one.''

\end{itemize}

Phil Gibbs is an ex-theoretical physicist who continues to do research in his spare time and also writes an entertaining blog \cite{Gibbs}. I recently got to know him because of our shared interests and I have learned a great deal from him: he is also a very accomplished mathematician.  However, he recently set up  a preprint archive on the internet, \url{http://vixra.org/},  catering to all who are unable to get their papers published in peer-reviewed journals or on the respectable archive, \url{http://arxiv.org/}, that professionals use (vixra =arxiv backwards). Now it is true that arxiv operates a filtering system. It is not perfect (they could perhaps make it more transparent), but on the whole I think they do a pretty good job. Readers can compare the quality of papers on arxiv and vixra and draw their own conclusions. In my opinion, vixra does serve a useful purpose: It eliminates once and for all the excuse that Grayling's ``unusual and eccentric minds'' can never get an airing.

\section{Repurposing string theory}
\label{repurpose}

In the forty years since its inception, string theory has undergone many changes of direction, in the light of new evidence and discovery:
\begin{itemize}
\item{1970s}
Strong nuclear interactions 
\item{1980s}
 Quantum gravity; ``theory of everything'' 
\item{1990s}
 AdS/CFT: QCD (revival of 1970s); quark-gluon plasmas 
\item{2000s}
 AdS/CFT: superconductors 
\item{2000s}
 Cosmic strings 
\item{2010s}
Fluid mechanics
\item{2010s}
 Black hole/qubit correspondence: entanglement in Quantum 
Information Theory 
 \end{itemize}
 
For example, by stacking a large number of branes on top of one another, Juan Maldacena \cite{Maldacena} showed that a (D+1)-dimensional spacetime with all its gravitational interactions, may be dual to a non-gravitational theory that resides on its D-dimensional boundary. If this so-called holographic picture is correct, our universe maybe like Plato' s cave and we are the shadows projected on its walls. Its technical name is the ADS/CFT correspondence. Maldacena's 1998 ADS/CFT paper has garnered an incredible 7000+ citations. Interestingly enough, this is partly because it has found applications outside the traditional ``theory of everything'' milieu that one normally associates with string and M-theory. These, frequently serendipitous, applications include quark-gluon plasmas, high temperature superconductors and fluid mechanics. 

String and M-theory have also had a profound influence on pure mathematics:
\begin{itemize}
\item{}
 ``Until some 20 years ago, modern physics and mathematics seemed to be set on different unrelated paths of discovery. Since then, however, their paths have been crossing with increasing frequency as theories in physics have enriched mathematics and as mathematical concepts have shaped and directed pathbreaking insights in physics."
 \end{itemize}
These are the words of Sir Michael Atiyah, one of the world's most renowned mathematicians and 1966 winner of the Fields Medal, the mathematician's equivalent of the Nobel Prize.  If further proof were needed, string theorist Edward Witten himself won the Fields medal in 1990.

ADS/CFT is not the only branch of string/M-theory that has found applications in different areas of physics. After all, as shown in the table, string theory was originally invented in the 1970s to explain the behaviour of protons, neutrons and pions under the influence of the strong nuclear force.

More recently, in 2006, the author showed that the Bekenstein-Hawking entropy of certain black holes arising in string theory is given by exactly the same mathematical formula that describes the entanglement of three qubits in Quantum Information Theory. This turned out to be just the tip of an iceberg and since then, with contributions from Peter Levay in Budapest, Renata Kallosh and Andrei Linde at Stanford and Sergio Ferrara at CERN, a two-way dictionary between black-hole phenomena and quantum entanglement has been drawn up. More recently the author and his graduate students Leron Borsten, Duminda Dahanayake, William Rubens at Imperial College teamed up with Alessio Marrani at CERN. We invoked this black hole-qubit/correspondence to predict a new result in quantum information theory. Noting that the classification of stringy black holes puts them in 31 different families, we predicted that four qubits can be entangled in 31 different ways \cite{Borsten}. (By the way, this particular aspect of the correspondence is not a guess or a conjecture but a consequence of the Kostant-Sekiguchi theorem). This can, in principle, be tested in the laboratory and we are urging our experimental colleagues to find ways of doing just that.

While accepting the utility of many of these discoveries, critics complain that they were not in the original string manifesto. They fail to acknowledge that none of them would have happened had their stop-string-theory campaign been successful.

In summary, serendipity in theoretical physics is the norm rather than the exception. Predicting where the next breakthrough will come from is well-nigh impossible, except to say that it will be from one of those places you least expected. So long as bright young people continue to find string and M-theory exciting these subjects should and will continue to thrive and yield useful results, whether or not a ``theory of everything'' is forthcoming. Provided, of course, that the powers that be (the research funding agencies, university deans and presidents and vice-chancellors and shapers of public opinion) take the trouble to see beyond the ill-informed criticism and act accordingly.

\section{Does it matter?}
\label{matter}

Here we return to the question:  if the majority of theoretical physicists were in favour of string theory, would it really matter if the general public or scientists in other fields were not? After all, many of my string theory colleagues argue that by answering such critics you are according them a dignity they have not earned and providing them with the oxygen of publicity\footnote{See the letter by Deser, Lawrence and Schnitzer \cite{deser} to Scientific American for a welcome exception.}. I beg to differ.

For example, Michel Andre, the ``Adviser to the Director General of the European Union responsible for research policy issues" writes:

\begin{itemize}
\item{Research EU, the magazine of the European research area \cite{Andre}:}

``The book by American physicist Lee Smolin, The Trouble with Physics,
is an all-out attack on string theory in theoretical physics. 

The most interesting feature of The Trouble with Physics, and the one that has
attracted the most attention, is his sociological analysis of the way
in which string theory has taken root in academic circles and the
mechanisms that allowed it to gain its present almost total dominance.
How can a community of like-minded scientists have secured such
a powerful position that it is now able to determine the course of
research, to monopolise public funding and to decide careers, to the
point of abolishing all alternative approaches? 

Smolin has all too readily been labelled a frustrated scientist bent
on revenge for his lack of personal recognition. While he clearly
empathises with the brilliant eccentrics at the fringes of the scientific
community, this does nothing to detract from the pertinence of Smolin's
ideas and observations.''
\end{itemize}
Shortly after this was written, funding for the two European research networks on string theory was withdrawn.  Coincidence?

There is other anecdotal evidence that the criticisms of string/M-theory of the kind outlined in the present article are taking their toll.

Here in the UK, bureaucrats at the The Engineering and Physical Sciences Research Council have been routinely ``office rejecting''\footnote{Without peer review and completely uncontaminated by any contact with a scientist.} grant proposals on string theory. Just recently, in fact, EPSRC completely abolished its Mathematical Physics portfolio. 

TV personality Lord Robert Winston is the Professor of Science and Society at my own university, Imperial College London, and sits on the council of the EPSRC. I have invited him to comment on the wisdom of these decisions.  Elsewhere, for example in the context of autism and MMR vaccines, he has criticised the undue influence of scientific mavericks supported by unqualified journalists (especially in the Daily Mail) in the face of mainstream scientific opinion:

\begin{itemize}
\item{The Independent \cite{Hughes}}

``Labour peer warns of `crisis' in science. The Labour peer Lord Winston warned last night of an international crisis in science, levelling the blame at protesters such as the fuel lobby, arts graduates, the press, and even fellow scientists.

He highlighted ``single issue protest groups" as a particular threat, accusing such groups of distorting public opinion via ``manufactured protest". ''
\end{itemize}

Sound familiar?

 \section{Acknowledgments}
I am grateful to Phil Gibbs and Steven Weinberg for comments
on the manuscript. The opinions expressed in the article are entirely my
own. The author is supported in part by the STFC under rolling Grant No. ST/G000743/1. 

\newpage
\appendix
\section{FAQ}
\indent

Q. Is your falsifiable prediction of 31 ways to entangle four qubits really a test of string theory?
Some physicists, including Shelly Glashow (Nobel Laureate), Edward Witten (Fields Medalist) and Jim Gates (member of the President's Council of Advisors on Science and Technology), have been quoted as saying it isn't.

A. Shelly and Edward both told me later they were responding to the journalist's question of whether it is test of string theory as a ``theory of everything'', to which they gave the same answer I gave: ``No''. I said it is a test of string theory's ability to make statements about quantum information theory; in that particular paper we had nothing to say about particle physics, cosmology or a theory of everything. Jim's objection was that it is not, in any case, a test of string theory itself but of supergravity (which is conventionally regarded as the low energy limit of string theory but which could, in principle, be regarded as a theory in its own right). This is an interesting point that Jim and I are still debating, but one that was lost in the media coverage\footnote{Since falsifiability of string theory is the single issue of Peter Woit's ``single-issue protest group'', his blog has had a running commentary on the black-hole/qubit correspondence, which may be summarised as (1) It's wrong (2) It's trivial (3) Mathematicians thought of it first. These comments are in keeping with Woit's unerring gift for inaccuracy. For example, he wrongly credits me with having told author Ian McEwan about the Bagger-Lambert-Gustavsson model in M-theory, which he then proceeds to criticise \cite{Woit3}.
Commenting on a recent BBC documentary about claims of faster-than-light neutrinos, he refers to ``.. trademark hype from string theorist Mike Duff about how string theory could explain this'' \cite{Woit4}.  In fact I said that, although superluminal travel is in principle possible in the ``braneworld'' picture of string theory, in my opinion this was NOT the explanation for the claims \cite{Duffbbc}.}

Q. But an Imperial College press release carried a headline ``New study suggests researchers can now test the `theory of everything' ''. Isn't this an example of string theory hype your critics are complaining about?

A. That was admittedly unfortunate. Members of the Imperial media team I dealt with were very professional and sensitive to avoiding such hype. Consequently, I approved a version of the press release quoting me as saying:

\begin{itemize}
\item{}
 ``This will not be proof that string theory is the right ``theory of everything'' that is being sought by cosmologists and particle physicists. However, it will be very important to theoreticians because it will demonstrate whether or not string theory works, even if its application is in an unexpected and unrelated area of physics.''
\end{itemize}
The contradictory ``theory of everything'' headline was added by someone else in the media team, without my knowledge or consent.  Then all hell broke loose on the blogosphere.

Q.  Criticism of string theory is not confined to Lee Smolin, Peter Woit and the blogosphere. Nobel Laureates Shelly Glashow and Gerard 't Hooft, for example, have also raised objections. 

A. I have the greatest respect for both of them, and if they could come up with a better way of reconciling gravity and quantum mechanics, I would gladly give up string theory and work on theirs.  Actually 't Hooft seems to have softened his stance \cite{hooft}, saying ``.. it is now possible to describe at least some members of the black hole family using string theory with multidimensional membranes, called D-branes, added to it...String theory is just an instrument to do calculations in regions of a theory that are otherwise inaccessible.''
I take this as an endorsement of the view expressed in Section \ref{repurpose} that the whole enterprise is worthwhile whether or not  ``a theory of everything'' is forthcoming.



\subsection*{ADDED NOTE ON EPSRC December 2011}

The abolition of Mathematical Physics from the EPSRC portfolio met with resistance from the community. See for example the written evidence to Parliament from the Institute of Physics 

\url{http://www.publications.parliament.uk/pa/cm201012/cmselect/cmsctech/618/618vw26.htm}:
\begin{quote}
``Mathematical physics used to be an independent sub-theme supported by EPSRC, but has now disappeared from the new remit recently published on the EPSRC website. This change will exclude most of the areas of mathematical physics that EPSRC has supported in the past. This includes, but is not restricted to, the ending of support for areas that might also fall under the remit of STFC. This change in remit is in stark contrast to EPSRC’s landscape document where mathematical physics is the only listed sub-theme to get the top rating for international profile/standing.''
\end{quote}

As result of this pressure, due in large part to my colleague Chris Hull, EPSRC has restored Mathematical Physics as a sub-theme. 
However, whereas the old remit covered ``classical and quantum field theory, gauge theory, theory of gravity and string theory, 
general relativity'', these topics are not mentioned in the new remit, which comes with the warning that  ``Grants on the Web should not be considered as a guide to remit''. 

\subsubsection*{UPDATE ON EPSRC June 2012}

Worst fears have been confirmed.  Of the eleven subthemes in its Mathematical Sciences programme, Mathematical Physics is the only one to be ranked excellent in international profile/standing by the EPSRC. Yet in the recent ``Shaping Capability'' exercise it is the only  subtheme to be reduced, in disregard of EPSRC's own International Review of Mathematical Sciences,

 \url{http://www.epsrc.ac.uk/ourportfolio/researchareas/Pages/mathphys.aspx}

EPSRC has also changed the definition of Mathematical Physics from:

\begin{quote}
``Theoretical physics with a significant mathematical content'' 
\end{quote}
to 
\begin{quote}
``New mathematics for theoretical physics''.  
\end{quote}
When asked why the community was not consulted on this, a spokesperson responded  ``The remit of our mathematics programme has NOT changed. The Maths programme does not fund theoretical physics''. 

 \end{document}